# Single-shot dynamics of spin-orbit torque and spin transfer torque switching in three-terminal magnetic tunnel junctions

Eva Grimaldi[1], Viola Krizakova[1], Giacomo Sala[1], Farrukh Yasin[2], Sébastien Couet[2], Gouri Sankar Kar[2], Kevin Garello[2] and Pietro Gambardella[1]

[1]*Department of Materials, ETH Zurich, 8093 Zürich, Switzerland*
[2]*imec, Kapeldreef 75, 3001 Leuven, Belgium*

**Current-induced spin-transfer torques (STT) and spin-orbit torques (SOT) enable the electrical switching of magnetic tunnel junctions (MTJs) in nonvolatile magnetic random access memories. In order to develop faster memory devices, an improvement of the timescales underlying the current-driven magnetization dynamics is required. Here we report all-electrical time-resolved measurements of magnetization reversal driven by SOT in a three-terminal MTJ device. Single-shot measurements of the MTJ resistance during current injection reveal that SOT switching involves a stochastic two-step process consisting of a domain nucleation time and propagation time, which have different genesis, timescales, and statistical distributions compared to STT switching. We further show that the combination of SOT, STT, and voltage control of magnetic anisotropy (VCMA) leads to reproducible sub-ns switching with a spread of the cumulative switching time smaller than 0.2 ns. Our measurements unravel the combined impact of SOT, STT, and VCMA in determining the switching speed and efficiency of MTJ devices.**

Switching nanomagnets by current injection offers unparalleled scalability, as well as low power and high speed operation compared to control via external magnetic fields[1-3]. Spin-transfer torques (STT)[1,4] are presently employed in memory and spin logic applications[5,6] to control the state of magnetic tunnel junctions (MTJ) via an electric current passing through the reference and free magnetic layers, which allows also for efficient read-out of the MTJ through the tunnel magnetoresistance (TMR). Time-resolved studies of magnetization reversal in spin valve[7,8] and MTJ devices[9-12] have shown that STT enables switching on a timescale of 100 to 1 ns, depending on the driving current[13] and external field[14]. However, STT switching is characterized by nonreproducible dynamic paths and incubation times up to several tens of ns long, which limit the reliability and speed of the reversal process to about 10-20 ns, even when mitigation strategies based on large driving currents or noncollinear spin injection are employed[13,15,16].

These limitations may be overcome by magnetization reversal driven by spin-orbit torques (SOT)[3,17-19], which has been recently demonstrated in three-terminal MTJs with in-plane[20,21] as well as out-of-plane magnetization[22-25]. SOT switching combines an in-plane current injection geometry with charge-to-spin conversion due to the spin Hall effect and interfacial spin scattering[3]. Such a geometry decouples the write and read current paths, improving the MTJ endurance and operation speed by minimizing electrical stress of the tunnel barrier and allowing for tuning the barrier thickness for high TMR, fast read-out, and minimal read disturbances. Moreover, in devices with perpendicular magnetization, the injected spin current is orthogonal to the quiescent magnetization of the free layer, thus providing an "instant on" torque that is expected to minimize the switching incubation time[25-27].

Measurements of magnetic dots[27,28] and MTJs[23,29] performed after the injection of current pulses, as well as time-resolved stroboscopic measurements of magnetic dots performed using either pulsed x-ray[30] or laser probes[31,32], have succeeded in determining the probability and average speed of SOT-induced magnetization reversal. However, the stochastic nature of the transient dynamics as well as the actual speed of the individual switching events are not accessible in these experiments. Here we report real-time single-shot measurements of SOT switching in a three-terminal MTJ, which disclose the full reversal dynamics during current injection, including both deterministic and stochastic events. Contrary to previous reports, we show that SOT switching involves a finite incubation time up to several ns long, which is determined by the interplay of current-induced SOT, in-plane field, and Joule heating. By comparing the time-resolved magnetization reversal induced by SOT and STT in the same device, we evidence substantial differences in the dynamics, time scales, and efficiency of the two mechanisms. Finally, by combining SOT, STT, and voltage control of magnetic anisotropy (VCMA), we demonstrate sub-ns switching with unprecedented reproducibility and narrow distribution of the switching times.

**Devices and measurement scheme**

Our samples consist of three-terminal perpendicularly-magnetized MTJs with top-pinned CoFeB reference layer, MgO tunnel barrier, and a 0.9 nm-thick CoFeB free layer deposited on a 3.7 nm-thick $\beta$-phase W current line[25]. The MTJ pillars have a circular cross-section with a diameter of 80 nm (Fig. 1a,b). The orientation of the magnetization of the free layer, either up (+z) or down (-z), is detected through the changes of the TMR ($R_{MTJ}$), as shown in Fig. 1c for the minor loop recorded as a function of out-of-plane external field ($H_z$). The loop shows 100% remanence and sharp transitions between the parallel (P) and anti-parallel (AP) MTJ states, indicating full-reversal of the magnetization with field. In the following, the magnetization of the reference layer is pointing up, unless indicated otherwise.

Figure 1d shows a schematic of the three-terminal rf circuit used to bias the junction and the SOT current line, and measure $R_{MTJ}$ during switching. Two phase-matched rf pulses with 0.15 ns rise-time, variable length $\tau_p$, and amplitude $V_0$ and $V_1$ are injected into the feed lines of the bottom and top electrodes, biasing the MTJ at the driving voltages $V_{SOT}$ and $V_{STT}$ for SOT and STT switching, respectively. The resistance of the W-line and MTJ are $R_W = 309\,\Omega$ and $R_{MTJ} = 3.5\,k\Omega$ (P), $6.5\,k\Omega$ (AP), respectively. As $V_{SOT}$ itself generates a nonzero difference of electric potential between the top and bottom of the MTJ pillar, $V_{STT}$ is a function of $V_{SOT}$, which in our case is given by $V_{STT} \approx 1.09\,(x - 0.46)V_{SOT}$, where $x = V_1/V_0$ (see Methods and Supplementary Figure S1). Thus, for SOT switching, we apply two pulses $V_0$ and $V_1$ with the same polarity such that $V_{STT}$ is either zero ($x = 0.46$) or very small ($x = 0.63$), with negligible effects on the magnetization dynamics (see Methods and Supplementary Figure S2). A nonzero $V_{STT}$ is required for the time-resolved measurements in order to measure $R_{MTJ}$ during pulse injection. By varying $x$, we can study the entire bias range spanning from SOT to STT switching, revealing effects due to the interplay of the two torques and VCMA. To study the magnetization dynamics during pulse injection, we measure the voltage at the output electrode of the MTJ using a 20 GHz real-time oscilloscope. The measured signal results from the sum of the currents flowing in-plane ($V_{SOT}/R_W$) and out-of-plane ($V_{STT}/R_{MTJ}$). The output voltage $V_{sw}$, normalized to the voltage difference before and after switching, corresponds to the variation of the out-of-plane magnetization of the free layer from the initial state ($V_{sw} = 0$) to the final state ($V_{sw} = 1$)[11-13,33] (see Methods and Supplementary Figures S3 and S4).

**Switching polarity and after-pulse switching probability**

The polarity of SOT switching is determined by the relative alignment of the current with the static in-plane magnetic field ($H_x$). This field is required in order to break the up-down degeneracy of the damping-like SOT and achieve deterministic switching[3,17,30]. $H_x$ is a free parameter in our study that can be eventually integrated in the MTJ design[34]. In our samples, $V_{SOT} > 0$ favors the up state and $V_{SOT} < 0$ favors the down state of the free layer for $H_x > 0$, and vice versa for $H_x < 0$, as expected for the negative spin Hall angle of $\beta$-W [35]. For STT switching, the polarity of $V_{STT}$ sets the final state independently of $H_x$: $V_{STT} > 0$ favors the AP state, whereas $V_{STT} < 0$ favors the P state. When SOT and STT are combined, we keep $V_{STT}$ below the critical switching threshold and find that the final state is always the one promoted by SOT, which depends neither on the orientation of the reference layer nor on the sign of $V_{STT}$.

Figure 1e compares the after-pulse switching probability ($P_{sw}$) for SOT and STT measured on the same device as a function of $\tau_p$. The free layer has excellent switching properties under both SOT and STT bias; however, the two switching mechanisms have very distinct characteristics. For SOT switching, $P_{sw}$ is symmetric with respect to $\pm V_{SOT}$, whereas for STT switching $P_{sw}$ is strongly asymmetric with respect to $\pm V_{STT}$. The decrease of $P_{sw}$ at large $|V_{STT}|$ corresponds to conditions where the switching is no longer deterministic and affected by back-hopping, which is stronger for AP-P transitions[11,12]. Moreover, SOT ensures 100% switching for all values of $\tau_p$ down to 0.27 ns, whereas STT results in $P_{sw} < 1$ for $\tau_p < 10$ ns. The critical voltages ($V_c$) for SOT and STT switching, estimated at $P_{sw} = 0.5$, are about 0.4 V and 0.75 V at $\tau_p = 10$ ns, respectively. In the following, we focus on the switching dynamics at the minimum bias voltages required to achieve $P_{sw} = 1$ (measured up to 5000 trials), unless indicated otherwise.

**Time-resolved SOT and STT switching: average dynamics**

We first report time-resolved measurements of $V_{sw}$ averaged over multiple SOT switching events. Figure 2a shows $V_{sw}$ as a function of time, averaged over 5000 pulses with $\tau_p = 20$ ns at low field. A first striking observation is that, contrary to expectations[26,27,30-32], SOT switching occurs via a two-step process that involves a rather long incubation time of several ns, during which the free layer magnetization appears to be at rest, and a shorter time during which the magnetization fully reverses from the initial to the final state. The incubation time decreases substantially upon increasing either $|H_x|$ (Fig. 2b) or $V_{SOT}$ (compare Fig. 2b,c), which makes it possible to achieve sub-ns reversal with an extraordinarily narrow spread of the total switching time, comparable to the rising time of the current pulse (0.15 ns; see Fig. 2c and Supplementary Figure S5). These observations also show that the incubation time was likely too small to be detected in previous pump-probe experiments, which were carried out under relatively large currents and in-plane fields[30-32].

STT switching, performed on the same device, exhibits remarkably different dynamics and timescales compared to SOT. Figure 2d shows $V_{sw}$ averaged over 500 pulses with $\tau_p = 20$ ns. In this average measurement, $V_{sw}$ appears to change in a gradual way, evolving towards the final state over the entire duration of the pulse. This behavior is in striking contrast with the SOT-induced reversal, for which $V_{sw}$ remains quiescent and later increases within a few ns to full saturation, well before the pulse is completed. Increasing $H_x$ lowers the total STT switching time, but not as strongly as for SOT (Fig. 2e). The two switching modes are thus characterized by distinct time scales, even though both STT[11-13,36,37] and SOT[30] switching in perpendicular devices are known to occur via domain nucleation and propagation.

Our measurements also yield information on the critical energy $E_c(\tau_p) = V_c^2 \tau_p / R$ required to switch the free layer, where $R$ is either $R_W$ or $R_{MTJ}$. The large difference in the channel resistance for SOT and STT results in critical current densities of $2.0 \times 10^8$ A cm$^{-2}$ for SOT, and $2.3 \times 10^6$ A cm$^{-2}$ (P-AP) or $4.2 \times 10^6$ A cm$^{-2}$ (AP-P) for STT at $\tau_p = 10$ ns. Thus, even though $V_c$ is lower, $E_c$ is about an order of magnitude larger for SOT compared to STT for pulses longer than 5 ns. Reducing $\tau_p$ to 0.27 ns, however, leads to a minimum $E_c$ of 0.4 pJ for SOT switching, which is unattainable by STT (Fig. 2f). These measurements show that SOT switching requires an improvement of the critical current density in order to become as efficient as STT at long timescales, but they also demonstrate that SOT allows for significant energy gains compared to STT because shorter pulses can achieve switching. A detailed discussion of the figures of merit relevant for SOT magnetic random access memory devices is reported in the Supplementary Note 5.

**Real time dynamics of SOT and STT switching**

To gain insight into the stochastic versus deterministic aspects of the switching process, we performed single-shot measurements at different bias and fields, ranging from SOT- to STT-dominated switching. Figures 3a,b show the time traces of $V_{sw}$ measured during individual P-AP switching events at low and high in-plane field for $V_{SOT}$ = +453 mV (the minimum bias ensuring $P_{sw}$ = 1 in all cases). From left to right, these measurements show the evolution of SOT-driven switching with increasing contributions of the STT bias (from STT having a minor influence against switching at $V_{STT}$ = -227 mV, to STT-assisted switching at $V_{STT}$ = +266, +513 mV).

All the time traces indicate deterministic switching from the initial to the final state before completion of the pulse, with no discernible intermediate states. However, for almost pure SOT at low field, the magnetization reversal is characterized by significant trace-to-trace variations of the incubation and transition times. These variations, which are not discernible in pump-probe measurements[30-32], reveal the stochastic nature of the SOT reversal process. The ensuing fluctuations of the total switching time exceed by an order of magnitude predictions based on micromagnetic models in confined magnetic dots at finite temperature[38,39]. Importantly, very significant improvements of both the total time and the reproducibility of the reversal process are obtained by increasing either the STT bias or $H_x$, as shown in Fig. 3a,b. These improvements are all the more remarkable when compared with the time traces recorded for pure STT switching (Fig. 3c,d), which are characterized by slow transitions lasting several ns.

Comparing the statistical distributions of the incubation and transition times measured as a function of bias and field yields further insight into the advantages of combining SOT with STT. We define the incubation time $t_0$ and transition time $\Delta t$ by fitting each time trace with a linear ramp that approximates the two-step switching process (solid lines in Fig. 3). Figures 3e-j report the distributions of $t_0$ and $\Delta t$ determined from 1000 identical single-shot measurements. Starting from almost pure SOT at low field, which is characterized by broad incubation and transition time distributions with mean values $\bar{t}_0$ = 9 ± 2 ns and $\overline{\Delta t}$ = 2 ± 2 ns, respectively, we observe a strong reduction of $\bar{t}_0$ down to 2.0 ± 0.3 ns upon increasing $V_{STT}$ to +513 mV, accompanied by a narrowing of the transition times to $\overline{\Delta t}$ = 0.8 ± 0.3 ns (Fig. 3e,f). Increasing $H_x$ has also striking effects, leading to a drastic reduction of $\bar{t}_0$ to 0.7 ± 0.2 ns and a more peaked distribution of $\overline{\Delta t}$ around 0.4 ± 0.1 ns (Fig. 3h,i). In these conditions, we measure a total switching time $t_0 + \Delta t$ with a mean of 0.9 ns and a standard deviation of 0.16 ns (Fig. 3j). Incrementing the SOT bias also leads to a significant reduction of the incubation time and distribution widths (Supplementary Figure S8). Overall, we find that combining SOT with STT leads to standard deviations of $\bar{t}_0$ and $\overline{\Delta t}$ that are about an order of magnitude smaller compared to STT, even for strong overdrive conditions in noncollinear MTJs[15].

**Origin of the incubation and transition times**

The observation of a ns-long incubation time contrasts with the fact that the SOT is orthogonal to the magnetization of the free layer, leading to an "instant-on" switching torque. Micromagnetic simulations, combined with finite element simulations of current-induced heat dissipation in the MTJ (see Methods), shed light on this apparent contradiction. First, the simulations confirm that SOT switching occurs in two steps, namely the nucleation of a reverse domain at one edge of the free layer at the incubation time $t_0$, and the propagation of a domain wall towards the opposite edge during the transition time $\Delta t$. Second, the simulations reveal that the switching dynamics is strongly affected by the temporal evolution of the magnetic parameters of the free layer due to Joule heating (Fig. 4a). For $V_{SOT}$ close to $V_c$, the SOT is not sufficient to initiate the reversal of a domain at the beginning of the pulse. The progressive increase of the temperature of the free layer during the current pulse induces a decrease of the magnetic anisotropy energy, which eventually lowers the energy barrier for domain nucleation, initiating the switching. Only the simulated time traces that include the temperature rise closely reproduce the experimental curves, showing a finite incubation time followed by a fast transition (Fig. 4b,c). The decrease of anisotropy has the strongest influence on $t_0$. Additional simulations show that the decrease of the saturation magnetization and exchange stiffness have minor effects. Further, we find that $t_0$ scales with $H_x$ and $V_{SOT}$, in agreement with our measurements. The simulations thus demonstrate the origin of $t_0$ and $\Delta t$, highlighting the importance of thermal effects for realistic modelling of SOT-induced switching.

## Separation of VCMA and STT effects on SOT switching

The results shown in the previous sections highlight the strong gains in switching speed as well as reduced distribution widths that can be achieved by simultaneously biasing the SOT and STT electrodes of a three-terminal MTJ. Yet, if the reversal is dominated by the SOT, what is really the effect of $V_{STT}$? The STT bias is known to induce two different effects, namely a spin transfer torque[1,4] and a decrease or increase of the magnetic anisotropy of the free layer via the VCMA[40,41]. Both STT[24,42] and VCMA[43,44] have been shown to assist SOT switching in three-terminal MTJs; however, their separate influence as well as their effect on the magnetization dynamics have not been studied so far.

Here we distinguish the two effects by noting that the STT depends on the orientation of the magnetization of the reference layer, whereas the VCMA does not. Therefore, by inverting the reference layer using an external field, we can investigate MTJ's configurations in which the STT alternatively favors or hinders switching for a given polarity of the VCMA. Figures 5a,b illustrate the case of $V_{STT} > 0$ for the two opposite orientations of the reference layer; the possible combinations of SOT, STT, and VCMA and their effects on switching are reported in Table I. We first consider P-AP reversal for the reference layer pointing up, such that STT is expected to assist (oppose) switching for $V_{STT} > 0$ ($< 0$). Figure 5c compares the averaged time traces obtained for negative and positive values of $V_{STT}$ at constant $V_{SOT}$. Clearly, $V_{STT} > 0$ accelerates the reversal relative to $V_{STT} = 0$, whereas $V_{STT} < 0$, apart from reducing the noise level due to the increased bias of the junction, has only a minor effect on the dynamics. This bias asymmetry is manifested also in the critical SOT voltage $V_c$ determined from the measurements of $P_{SW}$ as a function of $\tau_p$ (Fig. 5d), which show that $V_c$ reduces by up to 30% for $V_{STT} > 0$, whereas it does not change noticeably for $V_{STT} < 0$. This asymmetry is a first indication that STT is not the sole effect influencing the SOT-driven magnetization reversal.

To separate the effects of STT and VCMA, we plot in Fig. 5e the normalized critical voltages $v_c = V_c(V_{STT})/V_c(V_{STT} = 0)$ corresponding to the reference layer magnetization pointing up ($v_c^{\uparrow}$) and down ($v_c^{\downarrow}$) for different values of $V_{STT}$. For the shorter $\tau_p$, the main effect is a downward shift of both $v_c^{\uparrow}$ and $v_c^{\downarrow}$ for positive increments of $V_{STT}$, which is quantified by the average $\bar{v}_c = \left(v_c^{\uparrow} + v_c^{\downarrow}\right)/2$ (Fig. 5f). For the longer $\tau_p$, in addition to the shift of $\bar{v}_c$, we observe that $v_c^{\uparrow}$ becomes increasingly smaller than $v_c^{\downarrow}$, as reflected by the difference $\Delta v_c = \left(v_c^{\uparrow} - v_c^{\downarrow}\right)/2$ (Fig. 5g). As $\bar{v}_c$ and $\Delta v_c$ are, respectively, even and odd with respect to the inversion of the reference layer, we ascribe the first effect to the VCMA and the second to STT.

The shift of $\bar{v}_c$ is consistent with a reduction of the magnetic anisotropy induced by a positive bias of the free layer. By assuming that the perpendicular magnetic anisotropy of the free layer scales proportionally to $\bar{v}_c$, we estimate a VCMA coefficient of 57 fJ $V^{-1}$ $m^{-1}$ for positive $V_{STT}$ (see Supplementary Note 9), which is in agreement with previous measurements of the VCMA in CoFeB/MgO[44,45]. On the other hand, $\Delta v_c$ is consistent with a scenario in which the STT adds or subtracts to the SOT depending on the relative sign of $V_{STT}$ and reference layer magnetization (see Table I). $\Delta v_c$ increases with $V_{STT}$ and $\tau_p$, thus showing that STT-assisted switching becomes relevant at larger bias and longer timescales. A likely scenario is that the SOT initiates the switching, mainly assisted by VCMA in lowering the domain nucleation barrier, as suggested by the strong reduction of $\bar{t}_0$ (see Fig. 3e,h), and then drives the propagation of a domain wall across the free layer assisted by STT, which modifies the damping of the free layer and, consequently, the domain wall speed (see Fig. 3f,i).

**Conclusions**

Our results provide direct insight into the real-time dynamics of SOT-induced magnetization reversal, either alone or in combination with STT and VCMA. Close to the critical voltage, the SOT switching rate is limited by the time required to nucleate a reverse domain, which is strongly affected by Joule heating, whereas the transition time corresponding to domain expansion is comparatively short. By taking advantage of the multiple bias and magnetic configurations of a three-terminal MTJ, we demonstrate different strategies to achieve faster switching with unprecedented uniformity and narrow distribution of the individual switching events. The combination of SOT, VCMA, and STT is particularly promising for reducing the latency and jitter of the writing process in logic and memory applications requiring high speed and non-volatility.

## Acknowledgments

This work was funded by the Swiss National Science Foundation (Grant No. 200020-172775), ETH Zurich (Career Seed Grant SEED-14 16-2), and IMEC's Industrial Affiliation Program on STT-MRAM devices.

## Author contributions

E.G., K.G., and P.G. planned the experiments. S.C., F.Y., and K.G. designed and fabricated the samples. E.G. implemented the time-resolved electrical setup. E.G. and V.K. performed the measurements. G.S. and V.K. performed the micromagnetic simulations. E.G., V.K, K.G., and P.G. analyzed the results. E.G. and P.G. wrote the manuscript. All authors discussed the data and commented on the manuscript.

## Additional information

## Data availability

The data that support the plots within this paper and other findings of this study are available from the corresponding author upon reasonable request.

## Competing financial interests

The authors declare no competing financial interests.

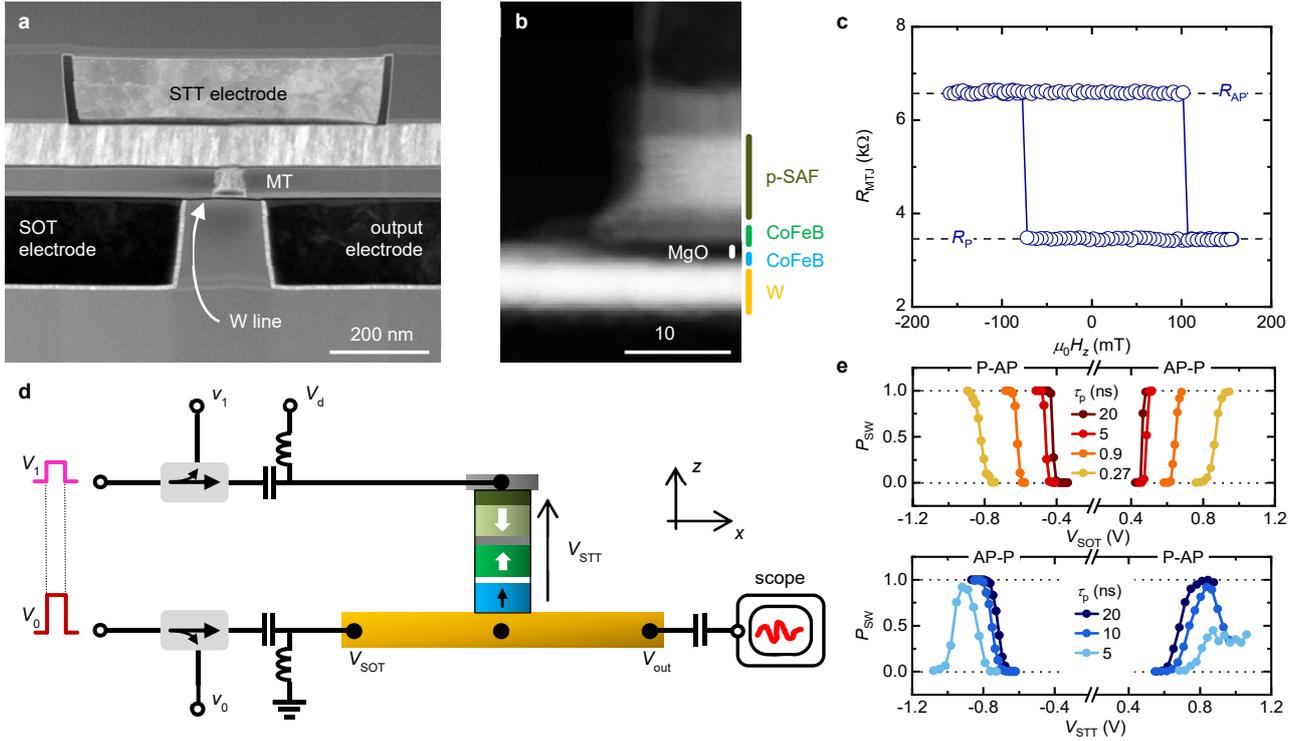

**Figure 1 | Schematic of the experimental setup and three-terminal MTJ. a**, Scanning electron microscope image of a three-terminal device with injection electrodes for SOT and STT switching. **b**, Detail of the MTJ pillar and W current line. **c**, Tunneling magnetoresistance as a function of external out-of-plane field. The 20 mT shift of the minor loop towards positive fields is due to the stray field of the top pinning layers, which favors the AP state. **d**, Electrical setup for the time-resolved measurements of the tunneling magnetoresistance during SOT and/or STT switching. **e**, Probability of SOT switching (top) at $\mu_0 H_x$ = -23 mT and STT switching (bottom) at $\mu_0 H_x$ = 0 mT as a function of the pulse amplitude for different pulse lengths. For each data point, $P_{sw}$ is calculated over 200 set-reset events by measuring $R_{MTJ}$ after each pulse.

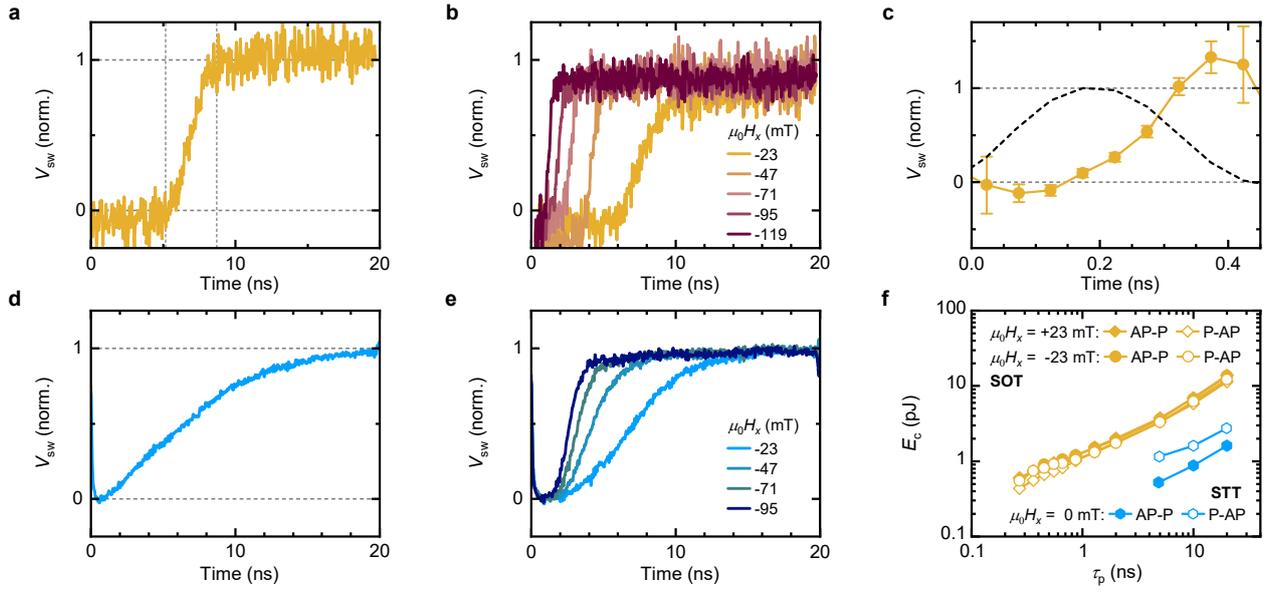

**Figure 2 | Average temporal evolution of the voltage signal during SOT and STT switching. a**, $V_{sw}$ due to AP-P magnetization reversal induced by SOT at $\mu_0 H_x = +23$ mT, $V_{SOT} = +475$ mV, and $\tau_p = 20$ ns averaged over 5000 pulses. The dashed lines indicate the region in which the free layer magnetization changes orientation during the multiple trials. **b**, Influence of $H_x$ on the P-AP reversal induced by SOT with $V_{SOT} = +453$ mV (5000 averages). **c**, Switching with sub-ns current pulses: $V_{sw}$ during AP-P magnetization reversal at $\mu_0 H_x = +23$ mT and $V_{SOT} = +997$ mV averaged over 1000 pulses with $\tau_p = 0.27$ ns and $V_{STT}/V_{SOT} = -0.5$. The data points are spaced by 50 ps, which corresponds to the sampling rate of our setup. The dashed line shows the measured pulse shape. The error bars have been calculated by adding in quadrature the noise of the switching and normalization voltage signals. **d**, $V_{sw}$ due to P-AP magnetization reversal induced by STT at $V_{STT} = +884$ mV, $\mu_0 H_x = 0$ mT and $\tau_p = 20$ ns averaged over 500 pulses. **e**, Influence of $H_x$ on the P-AP reversal induced by STT with $V_{STT} = +756$ mV (500 averages). The time traces in **d**,**e**, were offset prior to normalization in order to avoid division by close-to-zero values. In **a**,**b**,**d** and **e**, the pulses are 20 ns long and have the minimum amplitude required to achieve 100% switching at the lowest field. **f**, Comparison of the critical switching energy for SOT and STT as a function of $\tau_p$ measured in the lowest field conditions ($|\mu_0 H_x| = 23$ mT for SOT, $\mu_0 H_x = 0$ mT for STT).

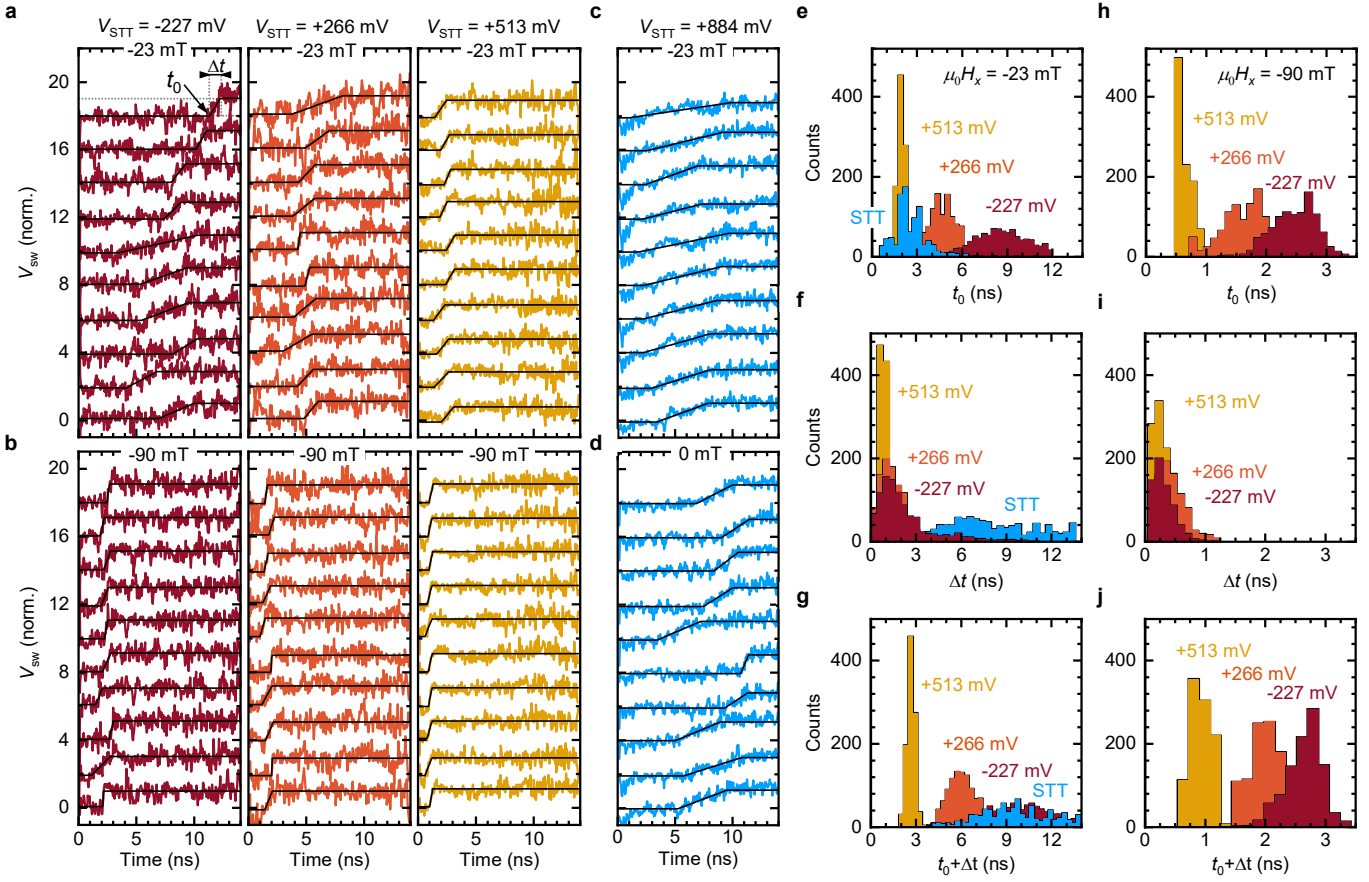

**Figure 3 | Single-shot measurements of SOT and STT switching. a,b,** Representative voltage time traces recorded during ten individual SOT-induced P-AP switching events induced by 15 ns long pulses with $V_{SOT} = +453$ mV and increasing STT bias $V_{STT} = -227, +266, +513$ mV at **a,** $\mu_0 H_x = -23$ mT and **b,** $\mu_0 H_x = -90$ mT. **c, d,** Representative voltage time traces recorded during ten individual STT-induced P-AP switching events at $V_{STT} = +884$ mV for **c,** $\mu_0 H_x = -23$ mT and **d,** $\mu_0 H_x = 0$ mT. The time traces have been vertically offset for clarity. Black solid lines are fits with a linear ramp used for extracting $t_0$ and $\Delta t$. **e,** Statistical distributions of the incubation time $t_0$, **f,** the transition time $\Delta t$, and **g,** the total switching time $t_0 + \Delta t$ for SOT switching at different STT biases (red-yellow) and for STT switching (blue) with $\mu_0 H_x = -23$ mT. Statistical distributions of **h,** $t_0$, **i,** $\Delta t$, and **j,** $t_0 + \Delta t$ with $\mu_0 H_x = -90$ mT. The histograms are obtained from the analysis of 1000 single-shot switching events. AP-P switching measurements are reported in Supplementary Figure S7.

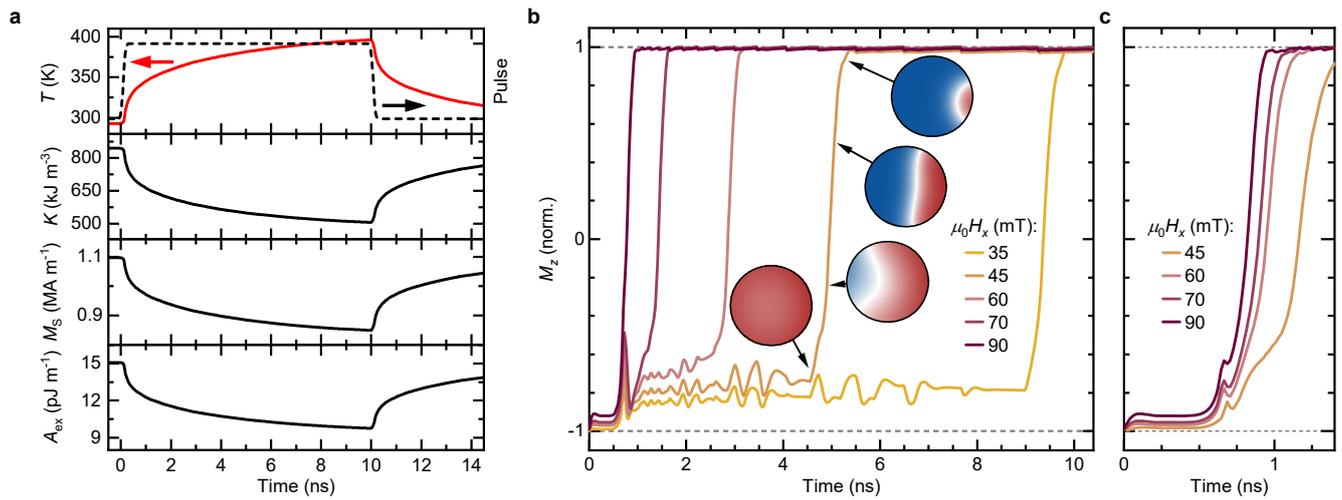

**Figure 4 | Temperature-induced variation of the magnetic parameters during pulse injection and micromagnetic simulations of SOT switching. a**, From top to bottom: variation of temperature, magnetic anisotropy energy, saturation magnetization, and exchange stiffness with time induced by a 10 ns current pulse (dashed line). **b,c**, Influence of $H_x$ on the magnetization reversal for **b**, 10 ns and **c**, 1 ns long pulses. The amplitude of the pulse is $1.10 \times 10^8$ A cm$^{-2}$ in **a,b**, and $1.22 \times 10^8$ A cm$^{-2}$ in **c**. The insets in **b**, are snapshots of the magnetic configuration at different times during magnetization reversal at $\mu_0 H_x = 45$ mT.

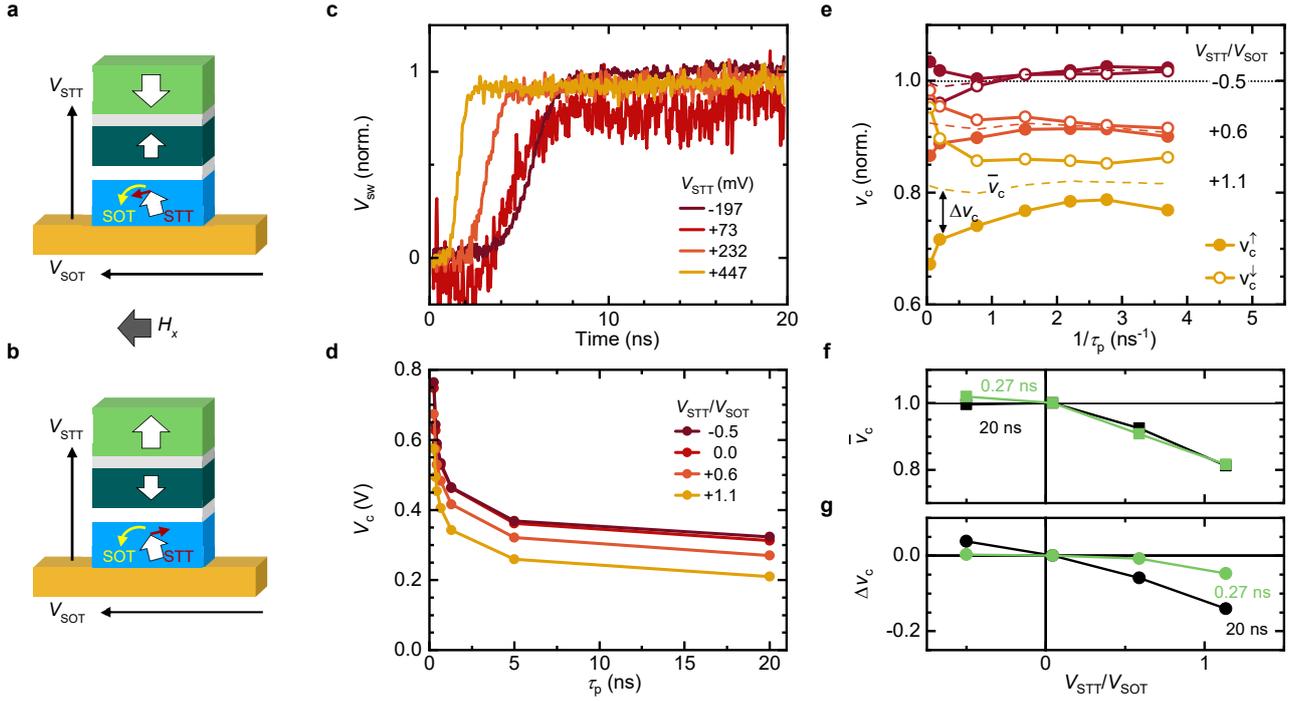

**Figure 5 | Combined effect of voltage control of magnetic anisotropy and STT on SOT-induced switching. a**, **b**, Schematics of SOT-induced switching for the reference layer pointing **a**, up and **b**, down. The black arrows indicate the bias, the white block arrows indicate the magnetization state. The yellow and red arrows schematize the switching directions favored by SOT and STT, respectively. In **a**, STT assists the SOT-induced switching, whereas in **b**, STT opposes the switching. **c**, Averaged voltage time traces of SOT-induced P-AP switching with different STT bias, $V_{STT} = -197, +73, +232, +447$ mV for $V_{SOT} = +394$ mV, and $\tau_p = 20$ ns. **d**, Evolution of the critical voltage as a function of pulse length for different $V_{STT}/V_{SOT}$ ratios. **e**, Normalized critical voltages $v_c^{\uparrow}$ and $v_c^{\downarrow}$ as a function of inverse pulse length for different $V_{STT}/V_{SOT}$ ratios. The magnetization of the reference layer points either up ($v_c^{\uparrow}$) or down ($v_c^{\downarrow}$), corresponding to the situations depicted in **a**, and **b**, respectively. The data are shown as a function of $1/\tau_p$ to better evidence the timescales at which STT and VCMA are relevant. **f**, Evolution of the average critical voltage $\bar{v}_c$ (VCMA-like contribution) and **g**, voltage asymmetry $\Delta v_c$ (STT-like contribution) as a function of $V_{STT}/V_{SOT}$ for short and long pulses. All measurements have been performed at $\mu_0 H_x = -90$ mT, $V_{SOT} > 0$ and correspond to P-AP (AP-P) switching for the reference layer magnetization pointing up (down) and free layer initially pointing up. Measurements performed for the free layer initially pointing down are reported in the Supplementary Figure S9.

| Reference layer | SOT switching | | STT switching | | SOT and STT | $\Delta v_\mathrm{c}$ | SOT and VCMA | $\bar{v}_\mathrm{c}$ |
|---|---|---|---|---|---|---|---|---|
| up | $V_\mathrm{SOT} > 0, H_x < 0$ | P-AP | $V_\mathrm{STT} > 0$ | P-AP | add | $< 0$ | add | $< 1$ |
| | | | $V_\mathrm{STT} < 0$ | AP-P | subtract | $> 0$ | subtract | $\approx 1$ |
| down | $V_\mathrm{SOT} > 0, H_x < 0$ | AP-P | $V_\mathrm{STT} > 0$ | P-AP | subtract | $> 0$ | add | $< 1$ |
| | | | $V_\mathrm{STT} < 0$ | AP-P | add | $< 0$ | subtract | $\approx 1$ |

**Table 1 | Switching of a three-terminal MTJ.** Magnetic configurations of the MTJ for up-to-down switching of the free layer induced by SOT and STT. The combined effect of STT and VCMA on SOT-induced switching are also shown, together with the observed changes of the normalized critical voltage asymmetry $\Delta v_c$ and average normalized critical voltage $\bar{v}_c$, respectively, relative to pure SOT switching. The entries in the table refer to $V_\mathrm{SOT} > 0, H_x < 0$, corresponding to the data reported in Fig. 5. A complete list of all the possible combinations of SOT, STT, and VCMA is given in the Supplementary Note 10.

**Methods**

**Sample fabrication.** The three terminal MTJ devices were fabricated on imec's 300 mm MRAM pilot line[33]. First, damascene W bottom electrode contacts (for the SOT and output electrodes) were patterned on a plasma-enhanced chemical vapor deposition $SiO_2$ substrate and planarized. The stack shown in Fig. 1a,b was then deposited *in-situ* at room temperature by physical vapor deposition in a 300 mm cluster tool (EC7800 Canon-Anelva). After deposition, the stack was annealed at 300 °C for 30 min in a magnetic field of 1 T. The MTJ consists of a perpendicularly magnetized top-pinned stack with composition W(3.7)/ $Co_{20}Fe_{60}B_{20}$(0.9)/ MgO ($RA$=14 $\Omega$ $\mu m^2$)/ $Co_{17.5}Fe_{52.5}B_{30}$(1.1)/ W(0.3)/ SAF(10.5) (thicknesses are in nm). The bottom layer is made of $\beta$-phase W with a resistivity of 140 $\mu\Omega$ cm and an effective spin Hall angle of -0.32 (Ref. 25); the combined resistivity of the W/$Co_{20}Fe_{60}B_{20}$ bilayer is 120 $\mu\Omega$ cm. The $Co_{17.5}Fe_{52.5}B_{30}$ reference layer is ferromagnetically coupled through a W spacer to a perpendicular synthetic antiferromagnet (SAF) consisting of Co(1.2)/ Ru(0.85)/ Co(0.6)/ Pt(0.8)/ [Co(0.3)/ Pt(0.8)]$_6$. The perpendicular anisotropy field and thermal stability factor of the free layer, $H_K = 270$ mT and $\Delta = 50$ k$_B$T, respectively, were estimated by measuring the switching probability $P_{sw}(H)$ as a function of external field $H$ applied parallel to the easy axis, and fitting it by:

$$P_{sw}(H) = 1 - \exp\left[\frac{-H_K f_0 \sqrt{\pi}}{2r\sqrt{\Delta}} \text{erfc}\left[\sqrt{\Delta}\left(1 - \frac{H}{H_K}\right)\right]\right], \tag{1}$$

where $r$ is the field sweep rate and $f_0$ the attempt frequency (here set to 1 GHz)[46].

The MTJ pillar was defined using 193 nm immersion lithography and ion beam etching at normal and grazing angle with specific stop conditions based on end-point detection to leave the W layer intact while patterning the pillar without producing sidewall shorts across the MgO barrier. The W layer was then patterned by lithography and ion-beam etched, forming a 220 nm-wide current line with resistance $R_W = 309$ $\Omega$. Finally, a dual damascene Cu top electrode (STT electrode) was fabricated to complete the electrical connection (Fig. 1a).

**Electrical setup.** The electrical setup consists of rf and dc paths, which are decoupled by bias-tees and a dc block, as seen in Fig. 1d. The dc path is used for after-pulse measurements of $R_{MTJ}$. The rf path is used to inject the rf pulses for current-induced switching and time-resolved measurements. Pulses of length $\tau_p = 0.27$-20 ns, defined as the full-width at half-maximum of the injected pulse, are generated by a single pulse generator with 0.15 ns rise-time. Each pulse is split by a power divider in two identical pulses that are then independently attenuated in order to adjust the ratio $x = V_1/V_0$ between the pulses applied to the STT and SOT feed lines. In the case of STT switching, the SOT terminal is disconnected from the output of the power divider and both ends are connected to ground through a 50 $\Omega$ resistance. A 20 dB directional coupler is placed on each injection line to dissipate the back-reflected pulses, while the coupled ports $v_0$ and $v_1$ are connected to the oscilloscope and used for triggering and calibration of the switching pulse amplitudes. The time-resolved signal $V_{out}$ is acquired on a third input of the oscilloscope. According to our coordinate system, $V_{SOT} > 0$ generates a current flowing along $+x$ and $V_{STT} > 0$ a current flowing along $-z$.

**Equivalent resistance setup**. The equivalent resistance setup of a three-terminal MTJ is presented in Extended Data Figure 1. The STT bias is given by $V_{STT} = v_t - v_b$, where $v_t$ and $v_b$ are the voltages applied to the top and bottom of the MTJ pillar, respectively. Kirchhoff's law then gives $\frac{v_b - V_{out}}{0.5\ R_w} = \frac{V_{SOT} - v_b}{0.5\ R_w} + \frac{V_{STT}}{R_{MTJ}}$. As $V_{out}$ is initially set to 0 V, we obtain $V_{STT} = \left(v_t - \frac{V_{SOT}}{2}\right)\Big/\left(1 + \frac{R_W}{4R_{MTJ}}\right)$. Assuming $R_W \ll R_{MTJ}$ leads to a simplified expression

$$V_{STT} \approx v_t - 0.5\ V_{SOT}. \tag{2}$$

Due to the impedance mismatch at the sample edge, a large fraction of each pulse $V_0$ and $V_1$ is back-reflected in phase. The resulting voltages at the STT and SOT electrodes are thus the sum of the incoming and reflected pulses, as probed by separate measurements with a tee inserted between the directional coupler and before the device, which give

$$v_t = 1.9\, V_1 \text{ and } V_{SOT} = 1.75\, V_0. \tag{3}$$

These measurements agree with the basic theory of impedance mismatched circuits, for which the applied voltage at the device under test (DUT) is $V_{applied} = \frac{2\, R_{DUT}}{R_{DUT} + 50\,\Omega} V_{in}$. At the SOT electrode, $R_{DUT} = R_W = 309\,\Omega$, gives $V_{SOT} = 1.72\, V_0$, whereas at the STT electrode, $R_{DUT} = R_{MTJ} = 3.5 - 6.5\,\text{k}\Omega$ gives $v_t \approx 1.97\, V_1$. Using Eqs. (2) and (3), the STT bias applied to the MTJ can be expressed as a function of $V_0$: $V_{STT} \approx 1.9\, V_1 - \frac{1.75}{2} V_0 \approx 1.9(\,x - 0.46)V_0$, or as a function of $V_{SOT}$:

$$V_{STT} \approx 1.09\, (x - 0.46)\, V_{SOT}. \tag{4}$$

In agreement with the last equation, the condition for SOT-switching without an STT bias, i.e., $V_{STT} = 0$ V, is $x = 0.46$. However, $V_{STT} \neq 0$ V is required to probe $R_{MTJ}$ during the time-resolved measurements. To verify that the effect of the bias on the switching is minor when $V_{STT}$ is small, we compared the after-pulse switching probability $P_{sw}$ when $x = 0.46$ ($V_{STT} = 0$ V) with $x = 0.63$ ($V_{STT} \approx 50$ mV for $\tau_p > 5$ ns), which is used for time-resolved detection of the SOT switching. In Extended Data Figure 2, we show that the critical voltages $V_c$ at different $\tau_p$ are identical for the two configurations. This indicates that the effect of $V_{STT}$ on switching is negligible when $x = 0.63$, and hence validates the interpretation of these time-resolved measurements as being representative of SOT switching.

**Time-resolved measurement and data normalization.** To obtain a normalized time trace $V_{sw}$, we compare the signal acquired on the oscilloscope during switching ($V_{P-AP}$) to a non-switching reference time trace acquired when the magnetization remains in the final state ($V_{AP}$) or in the initial state ($V_P$) for the same driving pulse, similarly to time-resolved measurements of STT switching in 2-terminal devices[11,13,33]. For the measurement of $V_{AP}$, we initialize the magnetization in the final state and apply the same pulse and field conditions as used for switching. For the measurement of $V_P$, we initialize the magnetization in the initial state and reverse the direction of the in-plane field $H_x$ to prevent switching. This is not possible for pure STT switching, as the final state only depends on the pulse polarity. Therefore, for STT, we acquire $V_P$ with an out-of-plane field $H_z$ stronger than the coercive field of the free layer, forcing the magnetization to remain in the initial state. Typical conditions are $|\mu_0 H_z| = 300$ mT for P-AP and $|\mu_0 H_z| = 85$ mT for AP-P switching. The procedure used to obtain $V_{sw}$ for P-AP switching is summarized in Extended Data Figure 3, the AP-P switching is measured and treated in the same manner. Starting from the measured voltage time traces $V_{P-AP}$ (blue), $V_P$ (green) and $V_{AP}$ (red), we determine the switching signal ($V_{P-AP} - V_P$) and the reference signal ($V_{AP} - V_P$) by subtracting the initial state (here $V_P$) from the other two signals. $V_{sw} = (V_{P-AP} - V_P)/(V_{AP} - V_P)$ corresponds to the variation of the out-of-plane magnetization from the initial state ($V_{sw} = 0$) to the final state ($V_{sw} = 1$). The origin of the dynamics ($t = 0$) is deduced directly from the measurement as the maximum of the derivative of the measured signal. The measurements can be performed only within the pulse duration when ($V_{AP} - V_P$) is non-zero. The situation schematized in the figure corresponds to P-AP switching with $V_{SOT} > 0$, for which $V_{P-AP}$ and $V_{AP}$ are acquired with $H_x > 0$, and $V_P$ is acquired with $H_x < 0$. Examples of measured $V_P$, $V_{P-AP}$, $V_{AP-P}$ and $V_{AP}$ time traces are shown in Extended Data Figure 4 for both SOT and STT switching.

All measurements reported here have been performed on the same device at room temperature and are representative of other devices with MTJ diameter ranging from 80 to 200 nm. In total, 14 different devices have been measured.

**Micromagnetic simulations.** The current-induced magnetization reversal was modelled combining micromagnetic simulations with finite element simulations of the current and temperature distribution in a simplified magnetic stack. The current flow and temperature increase due to Joule heating were simulated using COMSOL in an MTJ pillar consisting of a CoFeB layer (1 nm) sandwiched between the W current line (3.7 nm) and the MgO tunnel barrier (1 nm), with layers on top of MgO treated as a unique element of Cu with 12 nm total thickness. The diameter of the MTJ was 80 nm and the W line had a rectangular shape with dimensions $(300 \times 200)$ nm$^2$. The substrate as well as the planarization material surrounding the MTJ pillar was SiO$_x$. The simulated temperature evolution versus time was used as an input to the object oriented micromagnetic framework (OOMMF) code[47], with included Dzyaloshinskii-Moriya interaction (DMI) extension module[48] and a custom-built implementation of the damping- and field-like SOT. In the micromagnetic simulations, the MTJ was further simplified to the single CoFeB free layer. We considered the temperature dependence of the magnetization parameters according to established models: the saturation magnetization varied as[49] $M_S(T) = M_{S0}\left(1 - \frac{T}{T_C}\right)^{\beta}$, where $M_{S0}$ and $T_C$ are the saturation magnetization at T = 0 K and the Curie temperature, respectively. Similarly, the magnetic anisotropy and exchange stiffness[50] were scaled as $K(T) = K_0\left(\frac{M_S(T)}{M_{S0}}\right)^{p}$ and $A_{ex}(T) = A_{ex0}\left(\frac{M_S(T)}{M_{S0}}\right)^{q}$, where $K_0$ and $A_{ex0}$ represent the anisotropy energy density and exchange stiffness at 0 K. $M_{S0}$, $K_0$, and $A_{ex0}$ were selected to match the room-temperature values of the quantities expected for the studied devices ($M_S = 1.1$ MA m$^{-1}$, $K = 845$ kJ m$^{-3}$, and $A_{ex} = 15$ pJ m$^{-1}$), whereas the Curie temperature and critical coefficients were taken from literature values for CoFeB/ MgO ($T_C = 750$ K, $\beta = 1$, $p = 2.5$, $q = 1.7$)[49]. We further note that the temperature effects in the simulation are deterministic, i.e., we did not consider fluctuations of the thermal bath. The SOT effective fields $\mu_0 H_{DL} = -60$ mT and $\mu_0 H_{FL} = -15$ mT, and the DMI strength $|D| = 0.15$ mJ m$^{-2}$ were kept constant. We further verified that the observed magnetization dynamics is not specific to the particular choice of these parameters. In the simulation, the CoFeB layer magnetization is initially saturated down and let relax for 0.5 ns after applying $H_x$. A current pulse with 0.2 ns rising/falling edge and $|j_{SOT}| = 1.10$ (1.22) $\times 10^8$ A cm$^{-2}$ amplitude is applied for 10 (1) ns. After the pulse end, the magnetization is let relax for 4 ns. The simulated time traces (Fig. 4b,c) represent the $z$-component of the magnetization computed in a $(4 \times 4 \times 1)$ nm$^3$ mesh, averaged over the layer and normalized by $M_S$.

# Supplementary information

## Table of contents

**Supplementary Note 1. Equivalent resistance setup and partition of applied voltage pulses**

The equivalent resistance setup of a three-terminal MTJ is presented in Fig. S1. The STT bias is given by $V_{STT} = v_t - v_b$, where $v_t$ and $v_b$ are the voltages applied to the top and bottom of the MTJ pillar, respectively. Kirchhoff's law then gives $\frac{v_b - V_{out}}{R_W/2} = \frac{V_{SOT} - v_b}{R_W/2} + \frac{V_{STT}}{R_{MTJ}}$. As $V_{out}$ is initially set to 0 V, we obtain $V_{STT} = \left(v_t - \frac{V_{SOT}}{2}\right) \Big/ \left(1 + \frac{R_W}{4R_{MTJ}}\right)$. For $R_W \ll R_{MTJ}$ we can write:

$$V_{STT} \approx v_t - 0.5\, V_{SOT}. \tag{1}$$

The ratio between the injected pulses $V_0$ and $V_1$ is fixed by using two different sets of attenuators, with $V_1 = x\, V_0$. Due to the impedance mismatch between the pulse injection line and the three-terminal MTJ, $V_0$ and $V_1$ are reflected at the sample edge. Compared to the incoming pulse, the reflection is in phase and approximately equal to $0.9\, V_1$ at the STT electrode and $0.75\, V_0$ at the SOT electrode, as probed by separate measurements with a tee inserted between the directional coupler and before the device. The voltage amplitude at the STT and SOT electrodes are thus the sum of the incoming and reflected pulses and equal to:

$$v_t = 1.9\, V_1, \tag{2}$$

$$V_{SOT} = 1.75\, V_0.$$

This estimate agrees with the basic theory of impedance mismatched circuits, for which the applied voltage at the device under test (DUT) is $V_{applied} = \frac{2\,R_{DUT}}{R_{DUT} + 50\,\Omega} V_{in}$. At the SOT electrode, $R_{DUT} = R_W = 309\ \Omega$, giving $V_{SOT} = 1.72\, V_0$, whereas at the STT electrode, $R_{DUT} = R_{MTJ} = 3.5 - 6.5\ \text{k}\Omega$, giving $v_t \approx 1.97\, V_1$.

Substituting (2) in (1) then gives the STT bias applied to the MTJ as a function of $V_0$:

$$V_{STT} \approx 1.9\, V_1 - \frac{1.75}{2} V_0 \approx 1.9 \left( x - \frac{1}{2}\frac{1.75}{1.9} \right) V_0 \approx 1.9\, (x - 0.46) V_0, \tag{3}$$

or as a function of $V_{SOT}$:

$$V_{STT} \approx 1.09\,(x - 0.46)\, V_{SOT}. \tag{4}$$

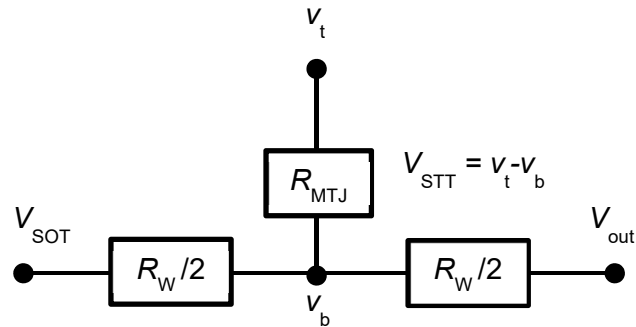

**Supplementary Figure S1 | Equivalent resistance setup of a three-terminal MTJ.**

**Supplementary Note 2. Validating the SOT-switching conditions for time-resolved measurements ($x =$ 0.63)**

The condition for SOT-switching without an STT bias is $x = 0.46$ (Supplementary Note 1). In this condition, $V_{\text{STT}} = 0$ V even upon the injection of an SOT pulse. However, $V_{\text{STT}} \neq 0$ V is required to probe $R_{\text{MTJ}}$ during the time-resolved measurements.

We compared the after-pulse switching probability $P_{\text{sw}}$ when $x = 0.46$ ($V_{\text{STT}} = 0$ V) with $x = 0.63$ ($V_{\text{STT}} \approx$ 50 mV for $\tau_{\text{p}} > 5$ ns), which is used for time-resolved detection of the SOT switching. In Figure S2, we show that the critical voltages $V_{\text{c}}$ at different $\tau_{\text{p}}$ are identical for the two configurations. This indicates that the effect of $V_{\text{STT}}$ on switching is negligible when $x = 0.63$, and hence validates the interpretation of these time-resolved measurements as being representative of SOT switching.

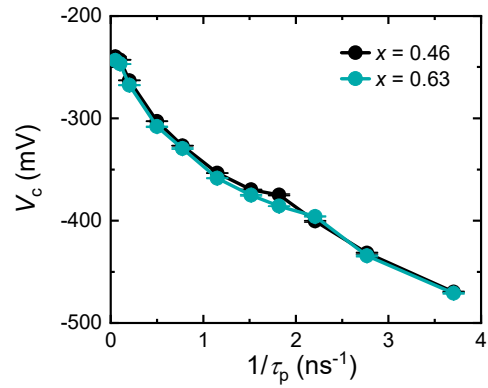

**Supplementary Figure S2 | Comparison of the critical voltage for zero and small STT bias.** The identical $V_{\text{c}}$ values and evolution with the pulse length $\tau_{\text{p}}$ for $x = 0.46$ and $x = 0.63$ indicate no noticeable effect of the small STT bias on the SOT switching.

**Supplementary Note 3. Measurement of time-resolved switching and data normalization**

For brevity, a detailed description of the time-resolved measurements and subsequent data treatment is provided for the P-AP switching only. The AP-P switching is measured and treated in the same manner.

To obtain a normalized time trace $V_{\text{sw}}$, we compare the signal acquired on the oscilloscope during switching ($V_{\text{P-AP}}$) to a non-switching reference time trace acquired when the magnetization remains in the final state ($V_{\text{AP}}$) or in the initial state ($V_{\text{P}}$) for the same driving pulse, see Fig. S3. For the measurement of $V_{\text{AP}}$, we initialize the magnetization in the final state and apply the same pulse and field conditions as used for switching. For the measurements of $V_{\text{P}}$, we initialize the magnetization into the initial state and reverse the direction of the in-plane field $H_x$ to prevent switching. This is not possible for pure STT switching, as the final state only depends on the pulse polarity. Therefore, for STT, we acquire $V_{\text{P}}$ under an out-of-plane field $H_z$ stronger than the coercive field of the free layer, forcing the magnetization to remain in the initial state. Typical conditions are $|\mu_0 H_z| = 300$ mT for P-AP and $|\mu_0 H_z| = 85$ mT for AP-P switching. Figure S3 summarizes the procedure used to obtain $V_{\text{sw}}$. Starting from the measured voltage time traces $V_{\text{P-AP}}$ (blue), $V_{\text{P}}$ (green) and $V_{\text{AP}}$ (red), we determine the switching signal ($V_{\text{P-AP}} - V_{\text{P}}$) and the reference signal ($V_{\text{AP}} - V_{\text{P}}$) by subtracting the initial state (here $V_{\text{P}}$) from the other two signals. $V_{\text{sw}} = (V_{\text{P-AP}} - V_{\text{P}})/(V_{\text{AP}} - V_{\text{P}})$ corresponds to the variation of the out-of-plane magnetization from the initial state ($V_{\text{sw}} = 0$) to the final state ($V_{\text{sw}} = 1$). The origin of the dynamics ($t = 0$) is deduced directly from the measurement as the maximum of the derivative of the measured signal. The measurements can be performed only within the pulse duration when ($V_{\text{AP}} - V_{\text{P}}$) is non-zero. The situation schematized in the figure corresponds to P-AP switching with $V_{\text{SOT}} > 0$, for which $V_{\text{P-AP}}$ and $V_{\text{AP}}$ are acquired with $H_x > 0$, and $V_{\text{P}}$ is acquired with $H_x < 0$.

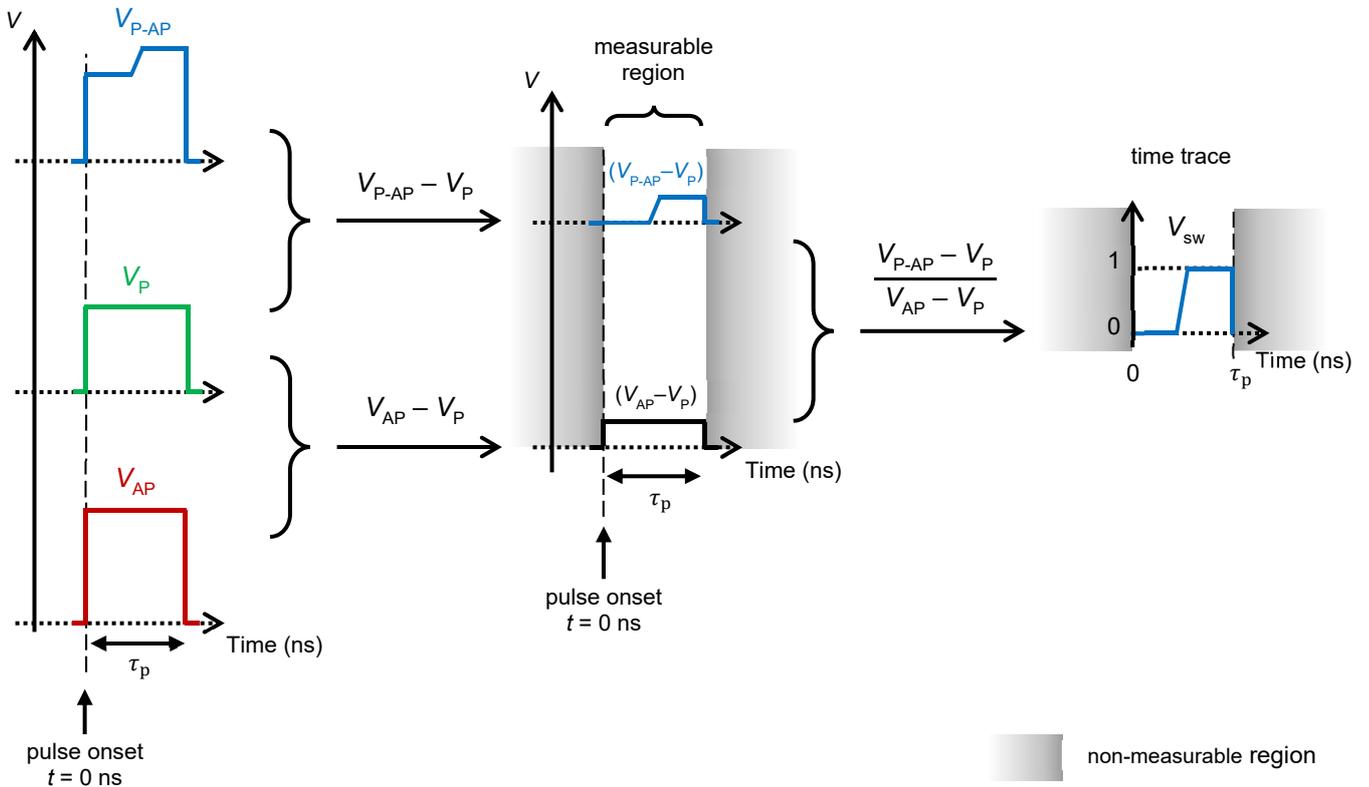

**Supplementary Figure S3 | Schematic of the acquisition and normalization protocol for P-AP switching.**

Examples of measured $V_P$, $V_{P-AP}$, $V_{AP-P}$ and $V_{AP}$ time traces are shown in Fig. S4 for both SOT and STT switching.

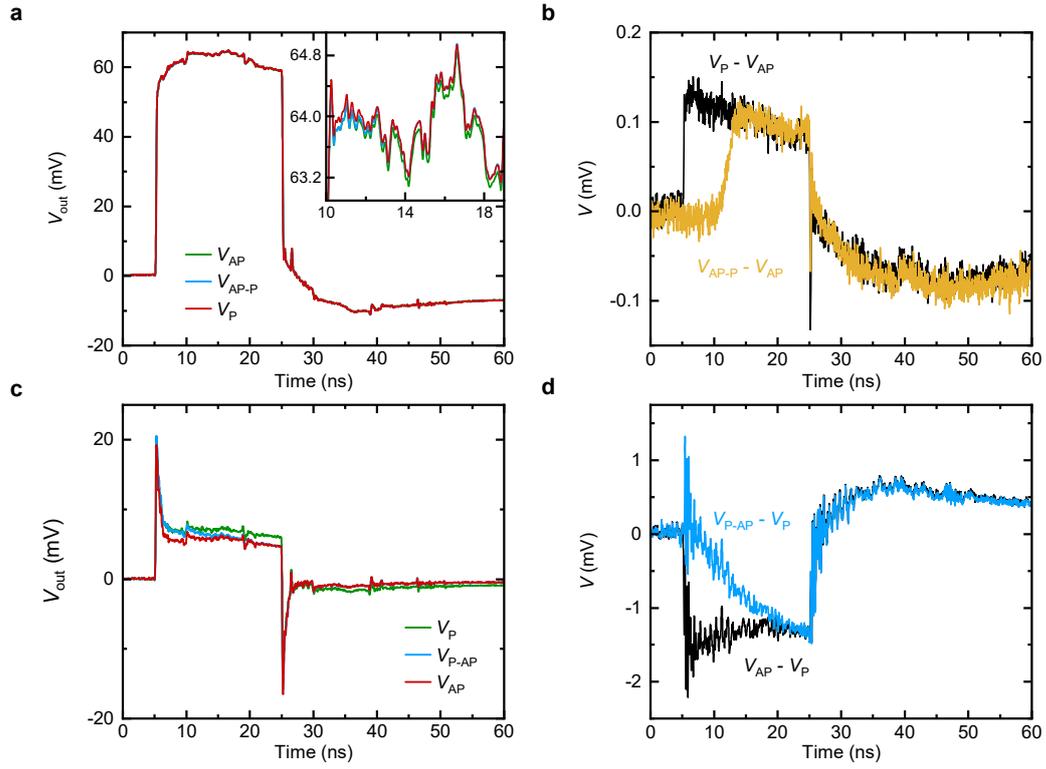

**Supplementary Figure S4 | Examples of $V_P$, $V_{P\text{-}AP}$, $V_{AP\text{-}P}$, and $V_{AP}$ time traces. a**, $V_{AP}$, $V_{AP-P}$ and $V_P$ for AP-P SOT switching. **b**, Corresponding switching and reference signals $V_{AP-P}-V_{AP}$ and $V_P-V_{AP}$, respectively. **c**, $V_P$, $V_{P-AP}$ and $V_{AP}$ for P-AP STT switching. **d**, Corresponding switching and reference signals $V_{P-AP}-V_P$ and $V_{AP}-V_P$, respectively. The traces are averages over 5000 (500) SOT (STT) switching events. These signals are used to obtain the normalized time traces shown in Fig. 2 a,b for SOT and STT switching, respectively.

**Supplementary Note 4. Comparison of the average switching times for SOT and STT**

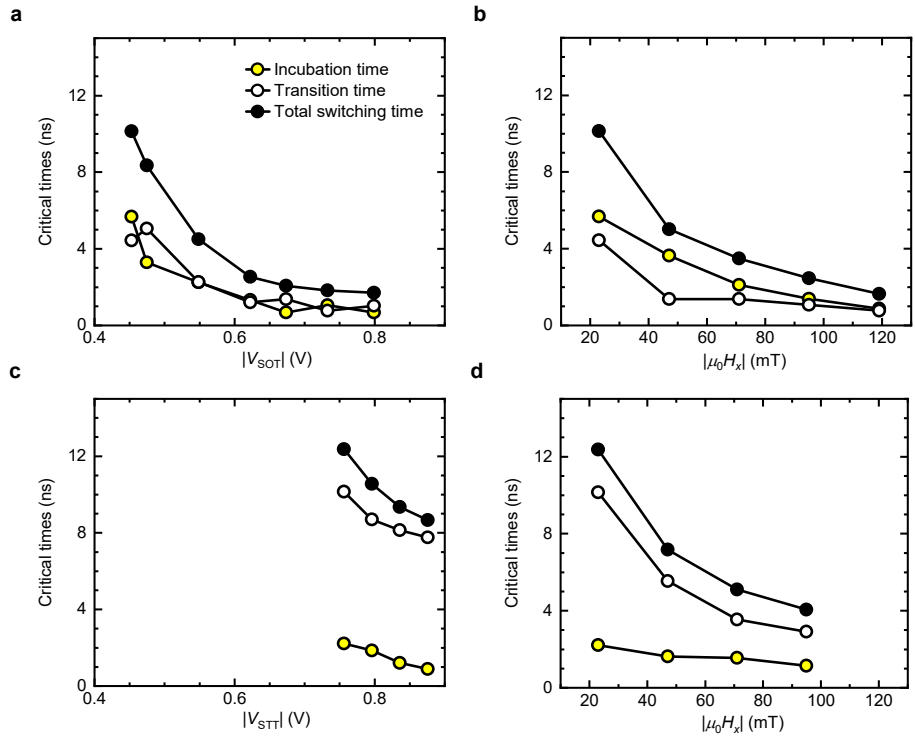

**Supplementary Figure S5 | Average SOT and STT switching times as a function of pulse amplitude and in-plane field. a,b,** Comparison of the average incubation time and transition time for P-AP SOT switching as a function of pulse amplitude ($V_{SOT} > 0$) at $\mu_0 H_x = $ -23 mT (**a**) and in-plane field $\mu_0 H_x$, at $V_{SOT} = $ +453 mV (**c**). **c,d** Comparison of the averaged incubation time and transition time for P-AP STT switching as a function of pulse amplitude ($V_{STT} > 0$) switching at $\mu_0 H_x = $ -23 mT (**c**) and in-plane field $\mu_0 H_x$, at $V_{STT} = $ +756 mV (**d**). In **b,d**, the pulse amplitude is the minimum allowing 100% switching ($P_{sw} = 1$) at $\mu_0 H_x = $ -23 mT.

**Supplementary Note 5. Figures of merit for SOT-MRAM devices**

Commercial deployment of SOT switching in MRAMs relies on meeting requirements that are specific to a given technology (CMOS node, memory array size, associated cell design and peripheral electronics) and end-user application. SOT-MRAM applications are primarily oriented towards replacing high-performance and high-density SRAM (8T and 6T families) at register, L1 and L2 level in CPUs and GPUs. The overall performances should not be considered only at the single cell level, but also at the array size level, which includes the dissipation and capacitances of the access lines as well as the control and sensing periphery. Therefore, proper benchmarking of a memory technology requires compact device models built from experimental data, bit-cell design optimization, and system-design power-performance-area-cost analysis. This analysis goes beyond the scope of this work. Nonetheless, we can discuss relevant aspects at the device level, which include: size, operation speed, endurance, power consumption, and CMOS compatibility.

*Cell size (cost).* The main difficulty today is that the work carried out on design is still limited, compared, e.g., to STT-MRAM. Preliminary analysis carried out at imec shows that standard SOT-MRAM designs are moderately competitive because of the too large number of controlling terminals. However, it is possible to reduce the number of terminals in an SOT bit cell by sharing some of them through smart designs, leading to a cell that can be 40% denser than SRAM cells. Below we compare cell size predictions for two different technology nodes based on imec's internal data:

- o Node = 16 nm / 5 nm
- o 8T-SRAM cell dimension: 0.1334 / 0.034 $\mu m^2$
- o 6T-SRAM (1:3:3) cell dimension: 0.124 / 0.26775 $\mu m^2$
- o SOT-MRAM simplest regular design: 0.162 / 0.028253 $\mu m^2$
- o SOT-MRAM optimized high density design: 0.0324 / 0.0162 $\mu m^2$

*Operation speed.* The read and write latencies (including delay time in access lines) should be < 1 ns for the typical target applications. We demonstrate here that the writing latency is largely matched. The reading latency is mostly dependent on the TMR (target >150%), reading current, resistance area product ($RA \sim$ 10 $\Omega \, \mu m^2$), and cell design (capacitance of the read bit line).

*Endurance.* Typical specification values are > $10^{14}$ read/write cycles. This is an area where SOT has clearly an advantage over STT, particularly at high speed. Various studies showed already endurance being tested > $10^{12}$ without cell degradation.

*Power consumption.* Write energy < 100 fJ per bit, ideally 25 fJ/bit. In the present devices, we have already reached 400 fJ by SOT (see Fig. S6a and Fig. 2f in the main manuscript). Reducing the device size and improving the VCMA and/or the SOT efficiency can bring very substantial reductions of the write energy. Read energy ~10 fJ per bit. The read energy is directly linked to the minimum current detectable by the periphery sense amplifiers and stand-by power. For fast reading with 100-150% TMR, the typical read current should be ~50 $\mu A$ in the low level state. Here SOT has a clear advantage over STT in that unintentional writing by the read current can be avoided. Further, compared to SRAM, the stand-by power of an SOT-MRAM is minimal.

*CMOS compatibility.* The compatibility with CMOS back-end-of-line processes is similar to STT-MRAM and already established[1]. Additionally, the size of the CMOS control transistors should be equal or smaller than the size of an SOT-MRAM cell in order to minimize footprint and be competitive with other technologies in terms of areal density. Therefore, a 3-terminal MTJ cell should match the power performances of a single CMOS transistor in terms of voltage and current delivery. Typically, a transistor in sub-28 nm nodes can deliver from 0.7 to 1.1 V and ~100 $\mu A$/fin. While the voltage is already within the specifications of the devices presented in this work, the writing current remains too large. Current values can be matched but would require more fins or larger planar transistors, increasing the cell size and cost.

Given the above, the main challenge in implementing commercial SOT-MRAMs lies in reducing the critical switching current while preserving the device functionality and speed. Strategies that allow for a reduction of the critical current are presented below.

The most straightforward approach is *downscaling*. Assuming a constant critical current density, an MTJ with a diameter of 30 nm placed on top of a 30 nm wide SOT-line would allow a critical current reduction from about 2 mA to 300 μA for 0.5 ns long pulses for SOT alone, that is, more than 80% with respect to the MTJs with 80 nm diameter and 220 nm wide SOT-line studied in this work (Fig. S6b, orange diamonds).

A second approach is to increase the *STT bias*. In this work, the bias range was restricted to $V_{STT}/V_{SOT} \leq +1.1$, which results in a decrease of the critical current of about 20% (Fig. S6b, open yellow diamonds). A stronger STT bias would lead to an even larger reduction of the critical current. Above a certain limit, the STT bias would lead to the emergence of STT-dominated field-free switching[2], which is observed around $V_{STT}/V_{SOT} = +4$ in our devices. At this point, however, the advantages of SOT-switching in terms of speed and endurance would be likely lost. For the scaled device, we estimate that a ratio of $V_{STT}/V_{SOT} = +3.6$ would be sufficient to reduce the critical current below 100 μA in the SOT-dominated regime (Fig. S6b, gray circles).

The third approach is to exploit the *VCMA effect*. Here we distinguish this effect from STT by averaging the critical current for P-AP and AP-P switching (see main text), which allows us to obtain the reduction of critical parameters due to VCMA alone. Based on our measurements (see, e.g., Fig. 5e), we estimate that the critical current would scale as $I_c = I_{c0} \times 45/(VCMA \, [\text{fJ V}^{-1} \text{ m}^{-1}])$, where $I_{c0}$ is the critical current at zero bias. Therefore, a VCMA coefficient of 120 fJ V⁻¹ m⁻¹ would be sufficient to achieve a critical current of 100 μA in the scaled device at 1 ns (Fig. S6b, black squares). Note that typical VCMA values reported in MgO-based MTJs reach up to 400 fJ V⁻¹ m⁻¹ (Refs. 3,4).

The fourth possibility is to increase the *SOT efficiency*, i.e., the effective spin Hall angle of the SOT line. In our devices, the SOT efficieny is about -0.32. Recent work has demonstrated values of 1 and above in transition-metal alloys, topological insulators, and oxide interfaces[4]. The critical current density is supposed to scale linearly with the SOT efficiency. However, further work needs to be done in order to assess the compatibility of these systems with device requirements (e.g., perpendicular magnetic anisotropy) and compatibility with CMOS and back-end-of-line processes.

We emphasize that measurements on real devices are needed in order to validate the strategies and extrapolations presented above. In particular when reducing the MTJ size, several factors can play a role in either decreasing the critical current (geometry, thermal stability, Oersted field) or increasing it (heat dissipation, transition to macrospin behavior). Notably, the flexibility of SOT enables to test and combine different strategies in order to meet the specifications for different memory classes.

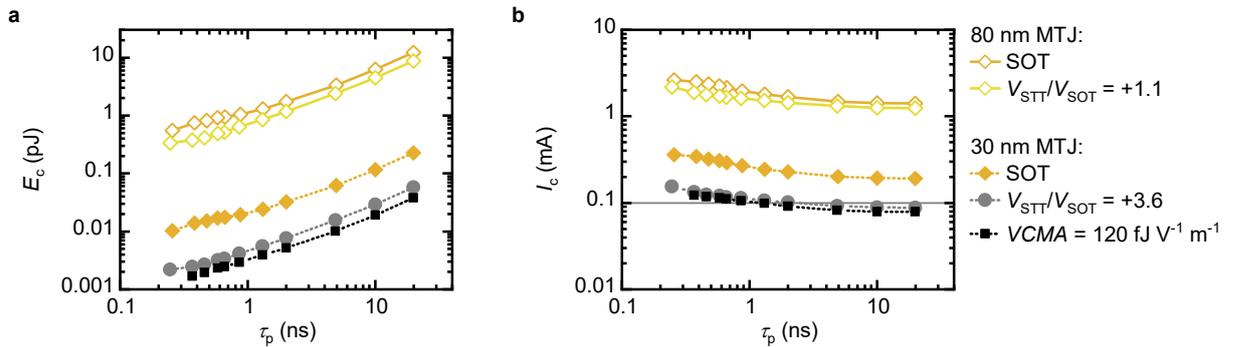

**Supplementary Figure S6 | Comparison of the critical switching energy and current in different conditions. a**, Critical energy $E_c$ and **b**, Critical current $I_c$ as a function of $\tau_p$. Open symbols correspond to P-AP reversal measured with $\mu_0 H_x$ = -23 mT for SOT switching (orange) and switching with $V_{STT}/V_{SOT} = +1.1$ (yellow). The full symbols represent an extrapolation of the experimental data to an MTJ with a diameter of 30 nm under various experimental conditions: pure SOT switching (orange diamonds), STT bias $V_{STT}/V_{SOT} = +3.6$ (gray circles), and VCMA of 120 fJ V⁻¹ m⁻¹ at $V_{STT}/V_{SOT} = +1.1$ (black squares). The 100 μA current limit is shown as a gray line in **b**.

## Supplementary Note 6. Single-shot SOT and STT switching in the AP-P case

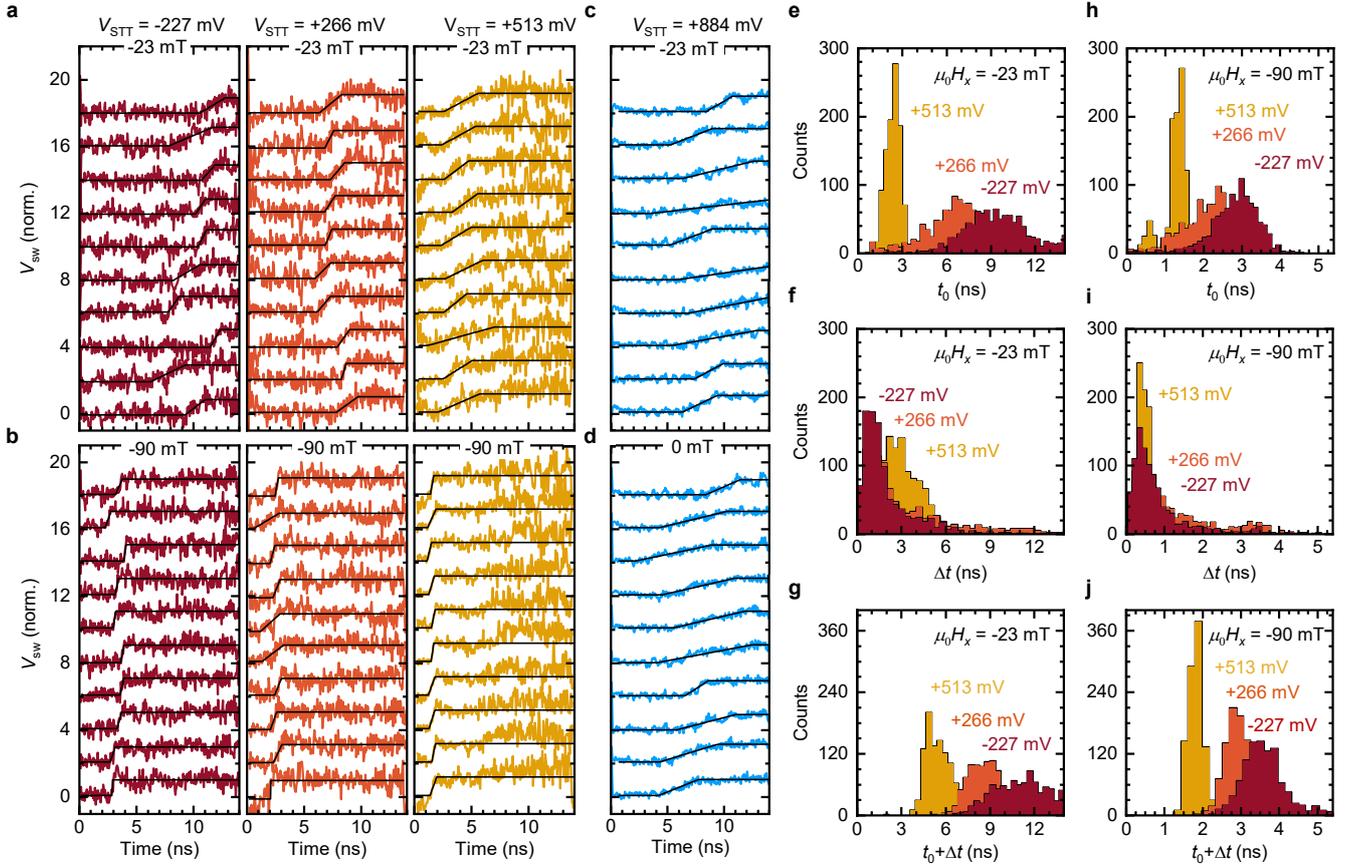

**Supplementary Figure S7 | Single-shot measurements of SOT and STT switching. a,b,** Representative time traces recorded during ten individual SOT-induced AP-P switching events induced by 15 ns long pulses with $V_{\mathrm{SOT}} = +453$ mV and increasing STT bias $V_{\mathrm{STT}} = -227$, $+266$, $+513$ mV at **a**, $\mu_0 H_x = +23$ mT and **b**, $\mu_0 H_x = +90$ mT. **c,d,** Representative time traces recorded during ten individual STT-induced AP-P switching events at $V_{\mathrm{STT}} = +884$ mV and **c**, $\mu_0 H_x = +23$ mT and **d**, $\mu_0 H_x = 0$ mT. The pulse amplitudes are the minimum ones required to achieve 100% switching in any of the shown configurations. The time traces have been vertically offset for clarity. Black solid lines are fits with a linear ramp used for extracting $t_0$ and $\Delta t$. **e,** Statistical distributions of the incubation time $t_0$, **f,** the transition time $\Delta t$ and **g,** the total switching time $t_0 + \Delta t$ at $\mu_0 H_x = +23$ mT for SOT switching at different STT biases. **h,** Statistical distributions of $t_0$, **i,** $\Delta t$ and **j,** $t_0 + \Delta t$ at $\mu_0 H_x = +90$ mT. The histograms are obtained from the analysis of 1000 single-shot switching events.

**Supplementary Note 7. Single-shot SOT switching for different $V_{SOT}$**

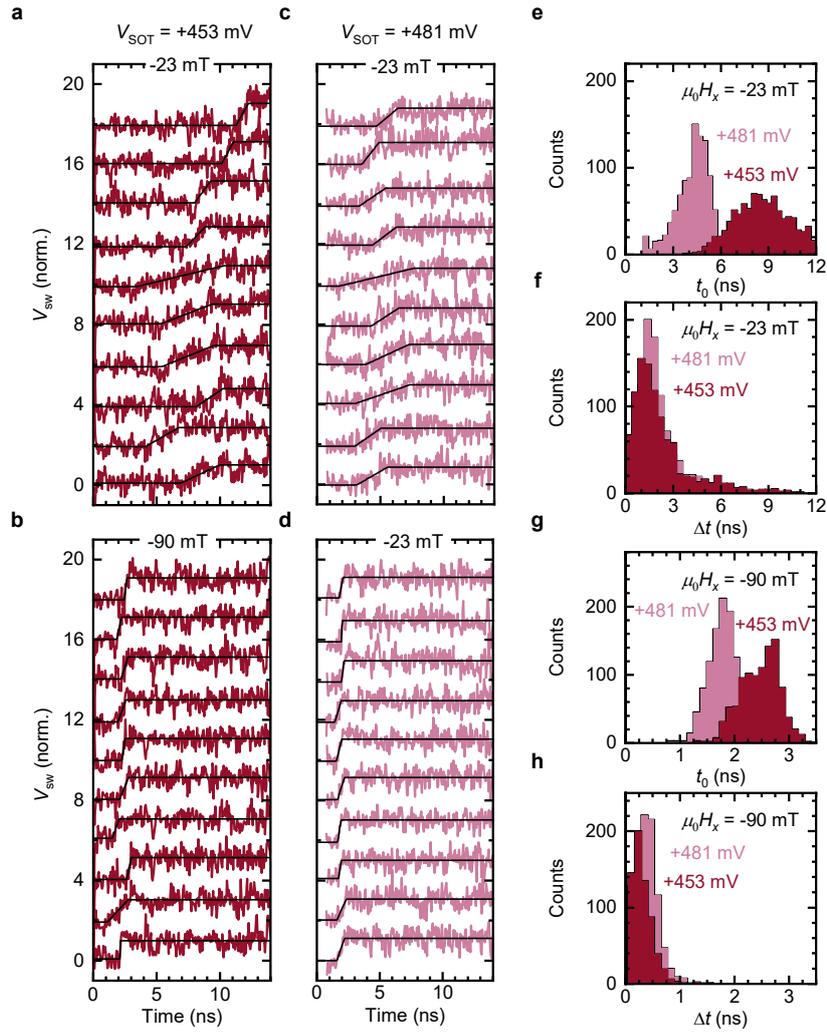

**Supplementary Figure S8 | Single-shot measurements of P-AP switching for different $V_{SOT}$ and $H_x$. a-d,** Representative time traces recorded during ten individual SOT-dominated P-AP switching events induced by 15 ns long pulses with **a,b,** $V_{SOT}$ = +453 mV, $V_{STT}$ = -227 mV and **c,d,** $V_{SOT}$ = +481 mV, $V_{STT}$ = -242 mV at **a,c,** $\mu_0 H_x$ = -23 mT and **c,d,** $\mu_0 H_x$ = -90 mT. The time traces have been offset for clarity. Black solid lines are fits with a linear ramp used for extracting $t_0$ and $\Delta t$. **e-h,** Distributions of **e,** $t_0$ and **f,** $\Delta t$ at $\mu_0 H_x$ = -23 mT, and **g,** $t_0$ and **h,** $\Delta t$ at $\mu_0 H_x$ = -90 mT. The histograms are obtained from the analysis of 1000 single-shot switching events.

**Supplementary Note 8. Effect of the VCMA and STT on SOT-induced switching for the free layer initially pointing down**

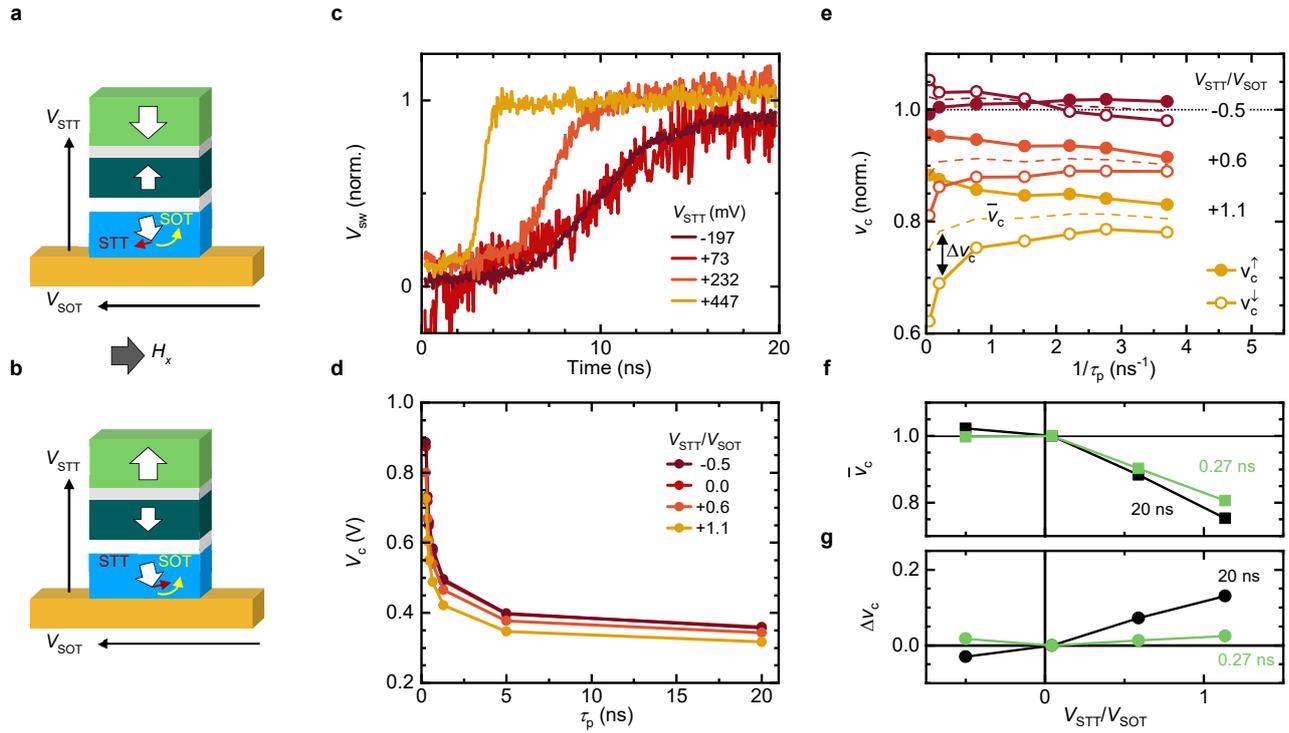

**Supplementary Figure S9 | Combined effect of VCMA and STT on SOT-induced switching. a,b,** Schematics of the SOT-dominated switching mechanism for the reference layer pointing **a,** up and **b,** down. Black arrows indicate the directions of rf currents, white arrows indicate the magnetization state. The STT (red arrow) given by $V_{STT}$ either **a,** opposes or **b,** assists the SOT switching induced by $V_{SOT}$ (yellow arrow). **c,** Averaged time traces of SOT-induced AP-P switching with $V_{STT}$ = -197, +73, +232, +447 mV for $V_{SOT}$ = +394 mV and $\tau_p$ = 20 ns. **d,** Evolution of the critical voltage as a function of pulse length for different STT contributions as given by the $V_{STT}/V_{SOT}$ ratio. **e,** Normalized critical voltages $v_c^\uparrow$ and $v_c^\downarrow$ as a function of inverse pulse length for different $V_{STT}/V_{SOT}$ ratios. The magnetization of the reference layer points either up ($v_c^\uparrow$) or down ($v_c^\downarrow$), corresponding to the situation depicted in **a,** and **b,** respectively. Evolution of **f,** the average cirtical voltage of $\bar{v}_c$ (VCMA-like contribution) and **g,** voltage asymmetry $\Delta v_c$ (STT-like contribution) as a function of $V_{STT}/V_{SOT}$ for short and long pulses. All measurements are performed at $\mu_0 H_x$ = +90 mT, $V_{SOT}$ > 0 and correspond to AP-P (respectively P-AP) switching for the reference layer pointing up (down) and free layer initially pointing down.

**Supplementary Note 9. Relationship between VCMA and critical switching voltage**

The VCMA effect was recently measured in MTJs[5] and results in an electric field dependent free layer effective anisotropy field ($H_{K,eff}$), characterized by the VCMA coefficient

$$VCMA = \frac{\mu_0 M_S t}{2} \frac{\partial H_{K,eff}}{\partial E}, \qquad (1)$$

where $\mu_0$ is the permeability of vacuum, $M_S$ the free layer saturation magnetization, $t$ the free layer thickness, and $E$ the applied electric field.

In general, the effective anisotropy field at a given $E$ can be linearized as

$$H_{K,eff}(E) = H_{K,eff}(0) + \frac{\partial H_{K,eff}}{\partial E} E. \qquad (2)$$

Due to the large resistance area product of the MgO layer with respect to the other metallic layers, we have $E \approx V_{STT}/t_{MgO}$, where $t_{MgO}$ is the thickness of the MgO barrier. We can thus write

$$H_{K,eff}(V_{STT}) = H_{K,eff}(0\text{ V}) + \frac{\partial H_{K,eff}}{\partial E} \frac{V_{STT}}{t_{MgO}}. \qquad (3)$$

By combining (1) and (3) we have

$$VCMA = \frac{\mu_0 M_S t}{2} \frac{t_{MgO}}{V_{STT}} \left( \frac{H_{K,eff}(V_{STT})}{H_{K,eff}(0)} - 1 \right) H_{K,eff}(0). \qquad (4)$$

Assuming that the critical voltage is proportional to the effective anisotropy field of the free layer,

$$V_c \propto \frac{e \mu_0 M_S t}{\hbar \theta_{SHE}} H_{K,eff}, \qquad (5)$$

where $e$ is the charge of the electron, $\mu_0$ the magnetic permeability of vacuum, $\hbar$ the reduced Planck constant, and $\theta_{SHE}$ the effective spin Hall angle of the W layer. Thus, assuming that the changes of the average of the normalized critical voltage are due to the VCMA effect, we have:

$$\bar{v}_c(V_{STT}) = \frac{V_c(V_{STT})}{V_c(V_{STT} = 0)} = \frac{H_{K,eff}(V_{STT})}{H_{K,eff}(V_{STT} = 0)}. \qquad (6)$$

From (4) and (6) we get

$$VCMA = \frac{\mu_0 M_S t\, t_{MgO} H_{K,eff}(0)}{2} \frac{\bar{v}_c(V_{STT}) - 1}{V_{STT}}. \qquad (7)$$

From the characterization of our sample we have $M_S = 900 \times 10^3\,\text{A m}^{-1}$, $t = 1$ nm, $H_{K,eff}(0) = 2.78 \times 10^5\,\text{A m}^{-1}$, and $t_{MgO} = 1$ nm. For $V_{STT} = 550$ mV, we have $\bar{v}_c(V_{STT}) = 0.8$, which gives a VCMA coefficient of 57 fJ V$^{-1}$ m$^{-1}$, in agreement with literature values[4,8].

**Supplementary Note 10. All possible combinations of SOT, STT, and VCMA for 3-terminal MTJ switching**

| SOT switching up to down | | | | STT switching | | SOT and STT | SOT and VCMA |
|---|---|---|---|---|---|---|---|
| REF up | $H_x > 0$ | $V_{SOT} > 0$ | AP-P | $V_{STT} > 0$ | P-AP | subtract | add |
| | | | | $V_{STT} < 0$ | AP-P | add | subtract |
| | | $V_{SOT} < 0$ | P-AP | $V_{STT} > 0$ | P-AP | add | add |
| | | | | $V_{STT} < 0$ | AP-P | subtract | subtract |
| | $H_x < 0$ | $V_{SOT} > 0$ | P-AP | $V_{STT} > 0$ | P-AP | add | add |
| | | | | $V_{STT} < 0$ | AP-P | subtract | subtract |
| | | $V_{SOT} < 0$ | AP-P | $V_{STT} > 0$ | P-AP | subtract | add |
| | | | | $V_{STT} < 0$ | AP-P | add | subtract |
| REF down | $H_x > 0$ | $V_{SOT} > 0$ | P-AP | $V_{STT} > 0$ | P-AP | add | add |
| | | | | $V_{STT} < 0$ | AP-P | subtract | subtract |
| | | $V_{SOT} < 0$ | AP-P | $V_{STT} > 0$ | P-AP | subtract | add |
| | | | | $V_{STT} < 0$ | AP-P | add | subtract |
| | $H_x < 0$ | $V_{SOT} > 0$ | AP-P | $V_{STT} > 0$ | P-AP | subtract | add |
| | | | | $V_{STT} < 0$ | AP-P | add | subtract |
| | | $V_{SOT} < 0$ | P-AP | $V_{STT} > 0$ | P-AP | add | add |
| | | | | $V_{STT} < 0$ | AP-P | subtract | subtract |

**Table S1: Combined effects of SOT, VCMA, and STT.** Magnetic configurations and corresponding combinations of STT and VCMA that either assist or hinder SOT switching in a 3-terminal MTJ based on a $\beta$-phase W SOT channel.